\documentstyle[epsfig]{mn2e}
\def\apj{ApJ}
\def\apjl{ApJ}
\def\msun{\rm {M}_{\odot}}

\def\simlt{\mathrel{\rlap{\lower 3pt\hbox{$\sim$}}\raise 2.0pt\hbox{$<$}}}
\def\simgt{\mathrel{\rlap{\lower 3pt\hbox{$\sim$}} \raise 2.0pt\hbox{$>$}}}
\def\lsim{\mathrel{\rlap{\lower 3pt\hbox{$\sim$}}\raise 2.0pt\hbox{$<$}}}
\def\gsim{\mathrel{\rlap{\lower 3pt\hbox{$\sim$}} \raise 2.0pt\hbox{$>$}}}

\def\msun{\rm {M}_{\odot}} 
\def\lta{\mathrel{\rlap{\lower 3pt\hbox{$=$}}\raise 2.0pt\hbox{$<$}}} 
\def\gta{\mathrel{\rlap{\lower 3pt\hbox{$=$}} \raise 2.0pt\hbox{$>$}}} 
\newcommand{\beq}{\begin{equation}}
\newcommand{\mnras}{MNRAS}
\newcommand{\aj}{AJ}
\newcommand{\aap}{A\&A}
\newcommand{\aapr}{A\&ARv}
\newcommand{\nat}{NATURE}
\newcommand{\physrep}{Phys. Rep.}
\newcommand{\eeq}{\end{equation}}

\begin{document} 
 
\title{High-redshift formation and evolution of central massive
  objects II: The census of BH seeds } %

\author[Devecchi, et al.]{B.~Devecchi$^1$, M.~Volonteri$^2$,
  E.~M.~Rossi$^1$, M.~Colpi$^3$, S.~Portegies Zwart$^1$ \\ $^1$ Leiden
  Observatory, Leiden University, P.O. Box 9513, 2300 RA Leiden, The
  Netherlands\\ $^2$ Astronomy Department, University of Michigan, 500
  Church Street, Ann Arbor, MI, 48109, USA \\ $^3$Dipartimento di
  Fisica G.~Occhialini, Universit\`a degli Studi di Milano Bicocca,
  Piazza della Scienza 3, 20126 Milano, Italy}

\maketitle \vspace {7cm}

\begin{abstract} 
We present results of simulations aimed at tracing the formation of
nuclear star clusters (NCs) and black hole (BH) seeds, in the
framework of the current $\Lambda$CDM cosmogony.  These BH seeds are
considered to be progenitors of the supermassive BHs that inhabit
today's galaxies.  We focus on two mechanisms for the formation of BHs
at high redshifts: as end-products of (1) Population III stars in
metal free halos, and of (2) runaway stellar collisions in metal poor
NCs. Our model tracks the chemical, radiative and mechanical feedback
of stars on the baryonic component of the evolving halos. This
procedure allows us to evaluate {\it when} and {\it where} the
conditions for BH formation are met, and to trace the emergence of BH
seeds arising from the dynamical channel, in a cosmological context.
BHs start to appear already at redshift $\sim$ 30 as remnants of
Population III stars. The efficiency of this mechanism begins
decreasing once feedbacks become increasingly important. Around
redshift $z\sim$ 15, BHs mostly form in the centre of mildly metal
enriched halos inside dense NCs. The seed BHs that form along the two
pathways have at birth a mass around $100-1000\,\msun$. The occupation
fraction of BHs is a function of both halo mass and mass growth rate:
at a given redshift, heavier and faster growing halos have a higher
chance to form a {\it native} BH, or to acquire an {\it inherited} BH
via merging of another system. With decreasing $z$, the probability of
finding a BH shifts toward progressively higher mass halo intervals.
This is due to the fact that, at later cosmic times, low mass systems
rarely form a seed, and already formed BHs are deposited into larger
mass systems due to hierarchical mergers. Our model predict that at
$z$=0, all halos above $10^{11}\,M_{\odot}$ should host a BH (in
agreement with observational results), most probably inherited during
their lifetime. Halos less massive then $10^9\, M_{\odot}$ have a
higher probability to host a native BH, but their occupation fraction
decreases below 10\%.
\end{abstract}


\section{Introduction} 

Supermassive BHs are currently thought to be at the heart of physical
mechanisms that shape the galaxy population we observe
today. Correlations of their masses with the large scale properties of
their galaxy hosts (Ferrarese \& Merritt 2000, Gebhardt et al. 2000,
G{\"u}ltekin et al. 2009, Tremaine \& Gebhardt 2009) are now accepted
as fundamental constrains for every theoretical model trying to
explain the evolution of the BH population.

The emergence of supermassive BHs of $10^9\,\msun$ already at redshift
6 (Fan et al. 2001, 2004) favours for an early formation and rapid
growth. Which process allows for the appearance of these seeds is
still not well established, nor what the characteristic mass of the
seeds is.  A range of possibilities have been explored that belong to
the following channels (see Volonteri 2010 for a more comprehensive
review):

\smallskip
\noindent
$\bullet$ Population III (PopIII) stars with masses above $260\,\msun$
are expected to end their lives leaving a relic BH of comparable mass,
because of negligible stellar mass loss (Heger et al. 2003, Madau \&
Rees 2001, Volonteri et al. 2003, Freese et al. 2008, Iocco et
al. 2008, Tanaka \& Haiman 2009 ). This small seed, housed in a
growing halo, later grows through accretion and mergers.

\smallskip
\noindent
$\bullet$ A compact star cluster can be subject to rapid segregation
of the most massive stars in its core. If mass segregation occurs
before copious mass loss, massive stars decouple dynamically from the
rest of the cluster and start colliding in a runaway fashion. The mass
spectrum evolves in such a way that a single very massive star (VMS)
grows quickly (Portegies Zwart et al. 1999). Its growth is terminated
once the reservoir of massive stars is exhausted, either via dynamical
collisions (as they are all engulfed in the VMS) or via stellar
evolution. At low metallicity (below $\approx 10^{-3}$ solar), stellar
mass loss is reduced compared to the solar metallicity case (see for
example Vink et al. 2008). A sufficiently massive VMS is then expected
to end its life leaving behind a remnant BH of a few hundred up to a
thousand solar masses (Yungelson et al. 2008, Belkus et al. 2007,
Portegies Zwart et al. 2004).

\smallskip
\noindent
$\bullet$ Metal free halos with virial temperatures above $10^4$ K can
be suitable sites for gas-dynamical instabilities leading to strong
gas inflows in the nucleus of the forming halo. A very massive star or
a quasi-star can form directly in these nuclei that collapses in the
form of a BH (Begelman et al. 2006, Bromm \& Loeb 2003, Dijkstra et
al. 2008, Eisenstein \& Loeb 1995, Haehnelt \& Rees 1993,Koushiappas
et al. 2004, Lodato \& Natarajan 2006, Volonteri \& Begelman 2010,
Spaans \& Silk 2006, Dotan et al. 2011 ).  Here, the BH seeds can be
as massive as $10^{4-5}\,\msun$ (Volonteri 2010).\footnote{It has been
  speculated that substantial inflows in major mergers can lead to
  massive BH seeds weighing more than $10^6\,\msun$ (Mayer et
  al. 2010). }

\smallskip
\noindent
$\bullet$ BH seeds can ensue as the result of processes arising in the very
early universe within a region of space where the gravitational force
overcomes pressure (Carr 2003, Khlopov et al. 2005) .

In Devecchi et al. (2010, Paper~I hereafter) and Devecchi \& Volonteri
(2009,  DV09 hereafter) we focused on the second channel. We explored
the formation of NCs and BH seeds in dark matter halos whose gas has
been polluted just above a critical metallicity $Z_{\rm crit}$ for
fragmentation and ordinary star formation. Gas, initially heated at
the virial temperature $T_{\rm vir}$ of the halo, cools down and forms
a disc.  Gravitationally unstable discs are subject to both mass
inflows and episodes of star formation, both in the central and outer
parts of the disc.  These early central star forming clusters can
provide scaled replica of the nuclear clusters (NCs) we observe today.
Our key finding was that these NCs are dense enough to be sites for
the onset of runaway collisions, ending with the formation of a BH
seed.

In this paper (Paper II) we include the BH population
rising from PopIII stars and study their joint evolution.  In order to
be effective, each of these mechanisms requires specific conditions to
be met. The hierarchical build-up of dark matter halos, together with
the evolution of their baryonic component, needs to be followed in
detail to constrain the efficiency of the two different BH formation
paths.  Critical factors that control and regulate this efficiency are
the chemical, radiative and mechanical feedbacks from the star forming
halos.  These feedbacks (neglected in Paper~I) affect the evolution of
the halo gas as they fix the available cooling channels and are then
fundamental ingredients if one wants to study the competition between
different BH formation paths.  We here implement these feedbacks
within a semi-analytical model for galaxy formation.  This is the
first time that this approach is used in the context of BH seed
formation.

The paper is organised as follows: in Section 2 we highlight the
procedure adopted in describing the evolution of the baryonic
component. We include the contribution of both PopIII and PopII-I
stars, with their chemical, mechanical and radiative feedback
effects. We describe how we populate metal free halos with PopIII
stars and briefly review the procedure adopted for halos with
$Z>Z_{\rm crit}$ (see Paper~I). In Section 3 we present our
simulations and in Section 4 we describe the results.  Section 5
contains our conclusions.

\section{Pop III formation}

\begin{figure*}\label{fig:bello}
\includegraphics[width=16cm]{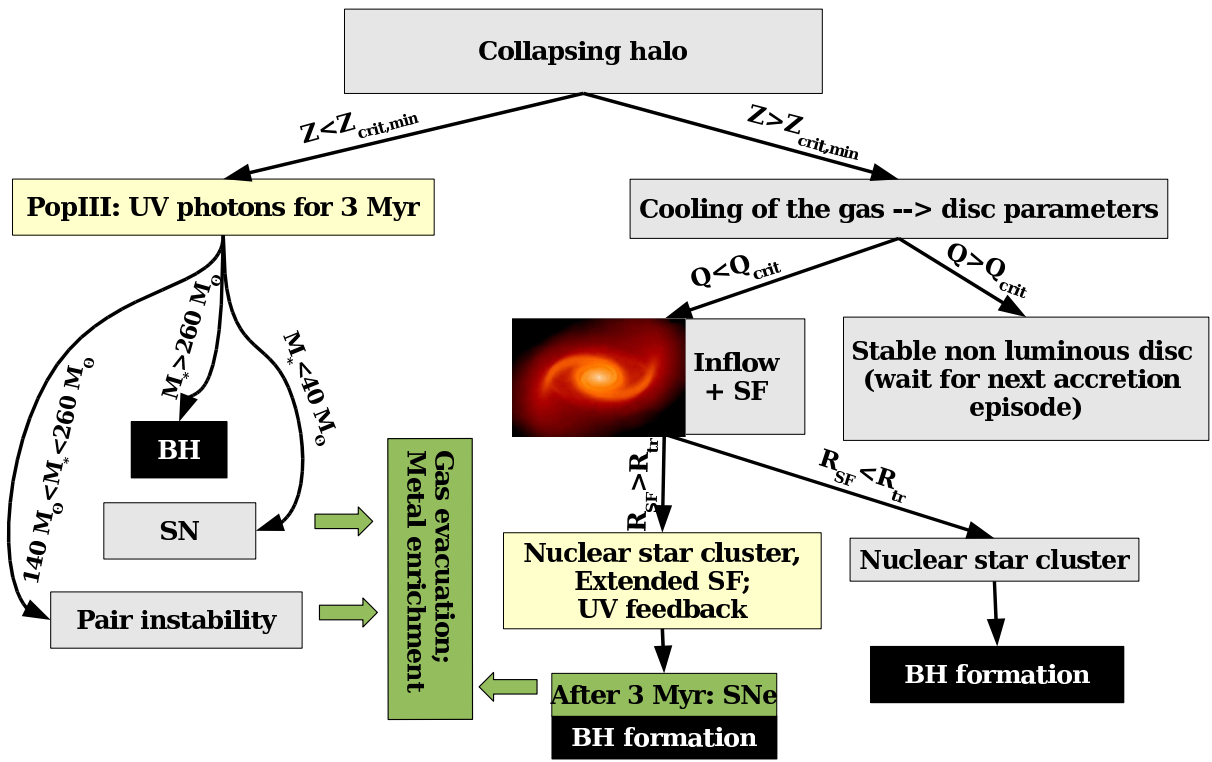}
\caption{Flow chart of the model (outlined in $\S$ 2) illustrating the
  different pathways that terminate with the formation of a seed BH,
  and the different prescriptions used to model the dissipative
  behaviour of the baryonic component. Dark grey areas (green areas in
  the on-line version of this manuscript) correspond to those part of
  the code associated with supernova feedback. Light grey boxes
  (yellow areas in the on-line version of this manuscript) correspond
  to those part of the code associated with radiative feedback. Black
  boxes indicate the presence of a channel for the possible formation
  of a BH seed (either as remnant of a PopIII star or as a result of
  dynamical instabilities in a newly formed nuclear stellar cluster).}
\end{figure*}

Here, we describe our recipe for the evolution of the baryonic
component hosted in a dark matter halo of mass $M_{\rm h}$, circular
velocity $V_{\rm h}$, virial radius $R_{\rm h}$ and spin parameter
$\lambda$. The behaviour of the gas strongly depends on the available
channels for cooling. These in turn rely upon the chemical species
available. It has been proposed that a critical metallicity $Z_{\rm
  crit}$ exists above which fragmentation into PopII-I star formation
occurs (Santoro \& Shull 2006, Bromm et al. 2001, Schneider et
al. 2006, Smith et al. 2009, Clark et al. 2009, Safranek-Shrader et
al. 2010). Hence, we distinguish between metal free halos (i.e., those
systems whose gas still has a pristine composition) and halos whose
gas has been enriched above $Z_{\rm crit}$. We assume that formation
of PopIII stars sets in only for $Z<Z_{\rm crit}$. Following DV09 and
Paper I, we assume a density dependent $Z_{\rm crit}$. The
relationships adopted in our study are those found by Santoro \& Shull
(2006). Our reference value for the minimum $Z_{\rm crit}\sim
10^{-5}\, Z_{\odot}$.

The first stars can form only in those halos where cooling allows
baryons to dissipate energy and condense in the centres of the dark
matter potential wells.  The first collapsing halos have virial
temperatures smaller than $10^4$ K, i.e. the temperature at which
cooling from electronic excitation of atomic hydrogen becomes
effective\footnote{In the following we will refer to mini
  (macro)-halos as those halos with virial temperature smaller
  (greater) than $10^4$ K.}. In order for their gas to cool down and
form the first stars, mini-halos with $T_{\rm vir}<10^4$ K must rely
on the less effective molecular hydrogen (${\rm H}_2$) cooling.

The critical minimum halo mass $M_{\rm PopIII,crit}$ that is necessary
in order for the gas to cool down efficiently and form a PopIII star,
roughly corresponds to $T_{\rm vir} \gsim 10^3$ K (Tegmark et
al. 1997). The presence of a UV flux impinging onto the halo increases
$M_{\rm PopIII, crit}$. We will briefly discuss this topic in the
following sub-section.

If enough ${\rm H}_2$ is present, gas can efficiently cool down to
$T\sim $ 200 K, reaching densities of $10^4$ cm$^{-3}$.  We here
assume that PopIII stars form in halos with $T_{\rm vir}\gsim 10^3$
K. We estimate the redshift at formation of the star following Trenti
et al. (2009). After the dark matter halo virialises at redshift
$z_{\rm vir}$, a time scale of the order of

\begin{eqnarray}
\tau_{\rm form}&=&7.6\times 10^{5}{\rm
  yr}\left(\frac{M_{\rm h}}{10^6\msun}\right)^{-2.627}\left(\frac{1+z_{\rm
    vir}}{31}\right)^{-6.94}\nonumber \\ & + & 8.82\times 10^{6}{\rm
  yr}\left(\frac{1+z_{\rm vir}}{31}\right)^{-3/2}
\end{eqnarray}
\noindent
is needed for the gas to cool down and collapse. 

Yoshida et al. (2003) develop cosmological simulations aimed at
studying PopIII star formation in a cosmological context. They show
that cooling of the gas can be prevented by dynamical heating of the
halo as a result of subsequent mergers. This delays PopIII star
formation in fast growing halos. We calculate the effect of dynamical
heating following Yoshida et al. (2003). We halt PopIII formation in
those halos where the dynamical heating rate is higher than the
cooling rate.

Despite the large number of studies performed, the initial mass
function (IMF) of PopIII stars is still poorly
constrained. Simulations of the initial phase of formation of these
object show that massive ($\sim10^3\,\msun$) clumps of gas can
collapse leading to the formation of a very dense, optically thick
core of $\approx 0.01\,\msun$. Gas in the envelope accretes into the
core, increasing the mass of the protostar. The characteristic final
mass at the end of the accretion process can be much less than the
initial clump mass: feedback effects can strongly reduce the ability
of the core to accrete material (see McKee \& Tan 2008, Omukai \&
Palla 2003). In addition, the characteristic mass of the star depends
on factors such as the presence of an external UV radiation
background, and/or the temperature of the CMB floor (see the
discussion in Trenti et al. 2009). We assume that a single PopIII star
forms in any given halo.  The mass of the star is extracted from a
distribution function $\Phi (m)\propto m^{-(1+x)}$ with $x=1.35$,
extending from a minimum and maximum masses of 10 and 300 $\msun$,
respectively (Omukai \& Palla 2003, McKee \& Tan 2008).  Note that
recent studies (see Glover 2008, Turk et al. 2009, Stacy et al. 2009,
Prieto et al. 2011) have pointed out that also PopIII stars can form
in clusters, where each of star's mass is much lower than the $\sim \,
100 \, M_{\odot}$ predicted by simulations that showed no
fragmentation and the formation of a single protostar. In Section 6 we
briefly discuss how this could affect our results.

\subsection{Radiative feedback}

Feedback effects from the first episodes of star formation can
strongly modify the efficiency at which gas is able to cool and
condense into stars.  We here try to obtain a rough estimate of the
radiative feedback on PopIII star formation efficiency, postponing the
discussion of chemical and mechanical feedback (via SN explosions) to
the next Section.

PopIII stars can produce copious amount of UV photons that can affect
the thermodynamics of gas inside the halo and in its neighbours.  In
particular, photons in the Lyman-Werner (LW) band (11.2-13.6 eV) can
photo-dissociate ${\rm H}_2$ molecules, thus suppressing cooling below
8000 K.  Formation of new zero metallicity stars is then suppressed
(or at least delayed) if the ${\rm H}_2$ dissociation rate is higher
than its formation rate.  Self-shielding of the gas in the densest
region can prevent ${\rm H}_2$ disruption by external radiation, this
effect being more efficient in the most massive halos. Machacek et
al. (2001) found a minimum threshold halo mass $M_{\rm TH}$ that is
necessary in order for Pop III star formation to set in (see also
O'Shea \& Norman 2008, Trenti et al. 2009, Wolcott-Green et
al. 2011). $M_{\rm TH}$ depends on the LW flux $J_{\rm LW}$ that
reaches the halo and can be written as

\begin{equation}
M_{\rm TH}/\msun=1.25\times 10^5 + 8.7\times 10^5 J^{0.47}_{\rm LW}.
\end{equation}
where $J_{\rm LW}$ is in units of $10^{-21}$ ergs$^{-1}$ cm$^{-2}$
HZ$^{-1}$.

For each metal free collapsing halo, we calculate the total LW flux
intercepted by the halo. Radiation sources are halos hosting an
emitting PopIII star and those with ongoing PopII/I star formation
(see next Section). We allow the new halo to form a PopIII star only
if its mass is above $M_{\rm TH}$.

Photon luminosities in the LW band are taken from Schraerer (2002) as a
function of the star's mass. The effective amount of photons that are
able to escape from the PopIII host halo depends on the details of the
system. Radiation can be trapped in the dense gas in which the PopIII
star is embedded. The escape fraction of photons in the LW band
$f^{\rm LW}_{\rm esc}$ depends on the luminosity of the star (and
consequently on its mass) and on the potential well of the host. A
critical mass of the host that corresponds to $f^{\rm LW}_{\rm esc}\sim
1\%$ has been calculated (Kitayama et al. 2004 but see also Alvarez et
al. 2006, Whalen et al. 2004)

\begin{equation}\label{eq:mfesc}
M^{\rm LW}_{\rm h,esc}=7.5\times 10^6 \left(\frac{m_{\rm
    PopIII}}{200\,\msun}\right)^{3/4}\left(\frac{1+z_{\rm
    vir}}{20}\right)^{-3/2}\,\msun.
\end{equation}

For halos with masses $M_{\rm h}>M^{\rm LW}_{\rm h,esc}$ we assume no
LW photons are emitted. On the other hand, we assume $f^{\rm LW}_{\rm
  esc}=1$ for $M_{\rm h}<M^{\rm LW}_{\rm h,esc}$. We adopt this sharp
transition motivated by the steep decrease of $f^{\rm LW}_{\rm esc}$
as a function of the host mass as shown in figures 4, 5, 6 in Kitayama
et al. 2004.

\subsection{Supernova feedback: metal enrichment and gas stripping}

After $\approx$ 3 Myr massive PopIII stars start to explode in
SN\ae. Metals processed in their centres are released into the halo gas
during the explosion and eventually propagate into the intergalactic
medium for sufficiently violent bursts. SN explosions in high
redshift halos can be extremely destructive. SNe driven bubbles can
push away part of the gas in the halo, eventually fully depriving the
host of its gas.

Metal yields and burst energies can strongly differ depending on the
mass of the progenitor star. The evolution of PopIII stars has been
studied in details by Heger \& Woosley (2002). They found that stars
with masses below $40\,\msun$ explode as SNe, releasing part of the
metals they produce.  Stars with masses between 140 and 260 $\msun$
are expected to be completely disrupted in pair instability SNe.
Stars with masses between 40-140 $\msun$ and above 260 $\msun$ instead
are expected to lead to direct BH formation with a mass comparable to
that of the progenitor star. For these two ranges of mass we assume no
metals are ejected. Each time a PopIII star explodes, a fraction of
its initial mass is converted into metals. Metal yields are taken from
Heger \& Woosley (2002). We assume the ejected metals efficiently mix
with the gas of the halo. The new metallicity of the halo gas is
calculated by taking into account for the total amount of metals
released, plus those already in place.

Explosion energy, $E_{\rm SN}$, of a PopIII star depends on its mass,
and its value ranges between $10^{51}-10^{53}$ erg (see Figure 1 in
Heger \& Woosley 2002). Values of $E_{\rm SN}$ as high as $10^{53}$
erg can easily evacuate a large fraction (up to 95\%) of the gas
initially present in the host (Whalen et al. 2008). The efficiency of
gas depletion depends on the depth of the potential well of the host
and on the ability of the star to photo-ionise and photo-evaporate the
gas. Mini-halos can be disrupted even by PopIII progenitors of 15
$\msun$ while more massive systems are much more resistant.  In order
to fix the amount of gas that remains in a halo, we here adopt the
same treatment as in Paper~I. We calculate the amount of energy
channelled into the outflow and assume the shell to evolve following
the Sedov-Taylor solution. This allows us to infer the energy of the
outflow, $K_{\rm sh}$, and its mass, $M_{\rm sh}$, at the moment it
reaches the virial radius. $M_{\rm sh}$ can be calculated by comparing
$K_{\rm sh}$ with the change in binding energy before and after the
explosion and it equals:

\begin{equation}
M_{\rm sh}=\frac{f_{\rm w}E_{\rm SN}}{2\left(1+\Omega_{\rm
    b}/\Omega_{\rm m}\right){GM_{\rm h}}/{R_{\rm h}}+({1}/{2})v^2_{\rm
    sh}}
\end{equation}
\noindent
where $f_{\rm w}$ is the fraction of the released energy channelled
into the outflow (for the exact expression of $f_{\rm w}$ see Paper~I
and Scannapieco et al. 2003) and $v_{\rm sh}$ the velocity of the
shell at $R_{\rm h}$ (see Appendix A of Paper~I for the details of the
calculation). This corresponds to a retention fraction, $f_{\rm
  ret}\equiv (M_{\rm gas}- M_{\rm sh})/M_{\rm gas}$.  After the
explosion we assume that a fraction $(1-f_{\rm ret})$ of the metal
yields produced in the PopIII star propagates into the medium
surrounding the halo.
 
\section{PopII-I star formation: the low-mass mode}

A single PopIII star explosion can easily pollute its host above the
critical metallicity for fragmentation $Z_{\rm crit}$. Polluted gas
that cools down settles in a disc, and can start forming stars in the
low mass mode. For the mass distribution of this stellar population we
adopt a Salpeter IMF, in the mass range 0.1-100 $\msun$. Gravitational
instabilities in these discs can also lead to mass inflow and nuclear
star formation so that a NC can form. Rapid dynamical evolution in the
cluster core may eventually lead to the formation of a BH seed. This
model for BH seed formation has been discussed in details in Paper~I
and in DV09. We here briefly review key points of the model.

1- Hot halo gas in virial equilibrium cools down at a rate
$\dot{M}_{\rm cool}$ computed according to the available cooling
channel (either molecular or atomic cooling, see Paper~I) and forms a
disc. The disc follows a Mestel profile and we calculate its
structural parameters as described in Paper~I.

2 - As gas cools down, the disc mass increases. The disc is unstable
if its Toomre parameter $Q$ (Toomre 1964) decreases below a critical
threshold $Q_c$.  Below $Q_c$ the disc develops non-axisymmetric
structures, leading to inflows and star formation. Inflows, at a rate:
\begin{equation}
 \dot{M}_{\rm
   grav}=\eta\left(\frac{Q^2_c}{Q^2}-1\right)\frac{c^3_s}{\pi G},
\end{equation} cause a redistribution of the disc mass that is transported
into the nuclear region. Here $c_s$ is the sound speed of the gas,
$\Sigma(R)$ its surface density profile and $\eta$ the inflow
efficiency. A steeper profile develops in the inner part of the disc
within a transition radius $R_{\rm tr}$.

3 - Stars form inside a star formation radius $R_{\rm SF}$. This is
calculated as that radius at which the adiabatic heating rate of the
gas equals its cooling rate. We assume that the star formation rate
follows a Kennicutt-Schmidt relation (Schmidt 1959, Kennicutt
1998). The star formation rate $\dot{M}_{*,\rm d}$ in the Mestel disc
scales with the disc parameters as
$\propto\left(\Sigma_0R_0\right)^{7/5}(R^{3/5}_{\rm SF}-R^{3/5}_{\rm
  tr})$. Star formation induces a reduction of the available gaseous
mass that can be transported in the nucleus. The net inflow rate is
$\dot{M}_{\rm inf}=\dot{M}_{\rm grav}-\dot{M}_{\rm *,d}$.

4 - Mass accumulated into the nucleus forms a compact cluster of
stars. We calculate the core collapse timescale $t_{\rm cc}$ of the
clusters and select those clusters with $t_{\rm cc}<3$ Myr, i.e.
systems in which the dynamical evolution precedes stellar
evolution. We follow the formalism of Portegies Zwart \& McMillan
(2002) to infer the mass $M_{\rm VMS}$ of the VMS as a function of the
cluster parameters. We select as possible sites for BH seed formation
only low metallicity clusters, i.e. systems with $Z<10^{-3}
Z_{\odot}$, as at higher metallicity the star looses mass in winds and
the final BH remnant has low mass (at most a few tens solar
masses). Each time $Z<10^{-3}Z_{\odot}$ and $M_{\rm VMS}>260\,\msun$,
i.e. it surpasses the threshold for pair instability SNe, a BH forms.

We also consider the contribution of PopII stars as sources of LW
photons. The LW luminosity associated to a star formation rate
$\dot{M}_*$ is calculated as $L_{\rm LW}=8\times 10^{27} \dot{M}_{\rm
  *}$ erg s$^{-1}$ Hz$^{-1}$ (Dijkstra et al. 2008). This luminosity
is associated to every halos that is forming PopII stars, given its
$\dot{M}_{\rm *}$. This information is used to compute the LW flux
impinging on surrounding halos when we determine whether H2 is
photo-dissociated or not (see for example Eq. 2).

SN explosions from evolved stars can lead to strong gas depletion,
particularly in the shallowest potential wells. We calculate the mass
loss rate from the halo $\dot{M}_{\rm sh}$ resulting from multiple explosion, as

\begin{equation}
\dot{M}_{\rm sh}=\frac{f_{\rm w}\nu_{\rm SN} E_{\rm SN}\dot{M}_{*}
}{2\left(1+\Omega_{\rm b}/\Omega_{\rm m}\right)GM_{\rm h}/R_{\rm
    h}+({1}/{2})v^2_{\rm sh}}
\end{equation}
\noindent
where $f_{\rm w}$ is the fraction of energy channelled into the
outflow, $\nu_{\rm SN}=0.00484$ is the number of SNe exploding after
the formation of a mass in star $M_*$ divided by $M_*$, $E_{\rm
  SN}=10^{51}$ erg is the energy of a single explosion and $v_{\rm
  sh}$ is the velocity of the outflow at the virial radius (see
Paper~I for details).

\section{Simulations}

Simulations are run in two steps. We first build up the dark matter
merger history, and we then populate this skeleton with galaxies,
including all the baryonic processes described in the previous
sections.

We keep track of the evolution of the dark matter component running
simulations with the code PINOCCHIO (Monaco et al. 2002a,b). PINOCCHIO
initialises a density perturbation field on a 3D grid. The density
field is evolved via the Lagrangian Perturbation Theory in order to
generate catalogues of properties (like mass, position and velocity)
and merger history of each virialised halo. 

We consider a cosmological volumes of 5 and 10 Mpc (co-moving) length.
We adopt a $\Lambda$CDM cosmology with $\Omega_{\rm b} $ = 0.041,
$\Omega_{\rm m}$ = 0.258, $\Omega_{\lambda}$ = 0.742, $h$ = 0.742, and
$n_s$ = 0.963, as given by five-year WMAP data (Dunkley et al. 2008).
Dark matter merger trees are followed up to $z=0$.

We follow the evolution of the baryonic component by applying the
prescriptions described in Paper I and in the previous Section. We
halt our simulations at redshift 6. After this redshift the BH
formation rate decreases, indicating that the bulk of the BH seed
population has already formed (see next Sections).

Figure 1 illustrates the recipes adopted in our semi-analytical model.
At first all halos are metal free with a baryonic gas fraction equal
to $\Omega_{\rm b}/\Omega_{\rm m}$. In each redshift interval and for
each halo, we apply the prescriptions described in Section 2 and
3. The evolution of each halo then depends not only on the merger
history of its dark matter component, but also on the earlier history
of its baryonic component. This is traced self-consistently starting
from the first PopIII star formation episode. For each halo, we
evaluate if and when a PopIII star forms. The star can either end
forming a BH seed or it can release metals: metal enrichment starts as
soon as halos are polluted by these first metals. Enriched gas starts
cooling again, at a rate depending on the amount of metals and gas
that the host is able to retain. 

Spatial positions are stored from the PINOCCHIO output with steps in
redshift of 0.25 (corresponding to 1-2 Myr at $z \sim 20$ and 10-20
Myr at $z \sim 6-10$).  This information is used in our semi-analytical
model in order to calculate radiative feedback from/on halos when the
emitting sources are active. Every time halo positions are updated we
also examine which sources are active.  Such fine time resolution is
required to estimate the effect of radiative feedback and the
production of LW photons, especially in the case of Pop~III stars.
Since the UV emission is limited to a few million years, the short
lifetime of these massive stars, the effect depends on the duration of
the emission and on the collapse time of halos surrounding each halo
hosting Pop~III stars.  If the time resolution is too coarse,
radiative feedback is not correctly evaluated. This is a consequence
of the short active time of some sources, particularly Pop~III stars;
if their activity is not monitored with a high enough time frequency,
the LW flux is initially underestimated. When this happens more halos
are allowed to form stars. This burst of star formation increases the
level of LW flux acting on the next redshift interval, so that in this
new timestep the amount of stars (and consequently LW photons) that
form, abruptly decreases. At $z>20$ when star formation is dominated
by Pop~III stars, our time-steps are shorter then the lifetime of
these objects. This guarantees that we can correctly account for the
presence of the emitting sources, and estimate the LW flux
adequately.

In the current cosmological scenario, merger events are quite frequent
in the history of dark matter halos. Every time two dark matter halos
merge we adopt the following prescriptions: we assume that the newly
formed halo has total, baryonic, stellar and metal masses given by the
sum of the progenitor ones. For mass ratio less than 1/10, we assume
the spin parameter of the main progenitor to be retained, otherwise a
new spin parameter is generated (according to the spin distribution
found in cosmological simulations; Bett et al. 2007). The properties
of the new baryonic disc are then calculated taking into account its
new mass and angular momentum. Its stability is evaluated and
eventually an episode of inflow and star formation sets in. Note that
in this way, star formation events are calculated taking into account
only for the stability of the structure at the end of the merger and
not for the tidal torques acting during the event. The metallicity of the
gas merger remnant is calculated adding together the amount of metals
of the two merging systems. Also in this case we assume efficient
mixing.

In some cases one or both merging halos host a black hole. Our
simulations are not able to follow the detailed evolution of the
merging process and to asses if the black hole hosted in the secondary
galaxy is able to reach the centre of the primary. Our prescription
only relies on the mass ratio between the two galaxies. For merging
systems with mass ratio greater than 1/10, we assume the secondary BH
to reach the centre of the merger remnant (Callegari et al. 2011). If
both galaxies originally hosted a black hole, the two are assumed to
merge. Below the 1/10 halo mass ratio the secondary BH (if present) is
left wandering in the new halo.

Our technique represents an optimal compromise between cosmological
simulations and standard semi-analytical codes. Our merger trees
contain the spatial and kinematical information on dark matter halos,
allowing us to determine the effects of feedback and metal pollution
on neighbouring halos, and taking into account mergers and dynamical
interactions. By using our semi-analytical model we can study gas
evolution and its impact on BH evolution at arbitrary resolution,
which is not possible even in the highest resolution cosmological
simulations (e.g., Sijacki et al. 2009, Bellovary et al. 2011)

\section{Black hole seed formation}

\subsection{Properties of the BH population}

\begin{figure}\label{fig:fbh}
\includegraphics[width=8cm]{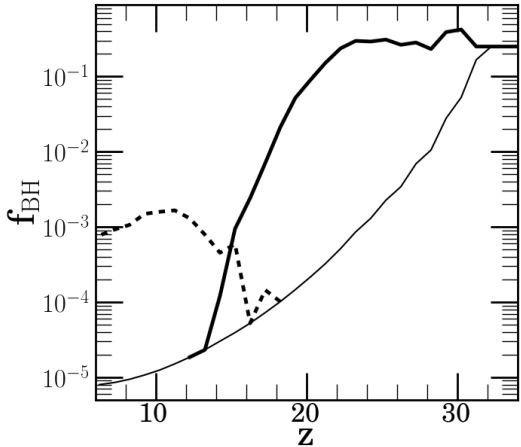}
\caption{Fraction of halos $f_{\rm BH}$ that at a given redshift $z$
  are forming a BH seed in situ, either from the PopIII (solid line)
  channel and the dynamical (dashed line).  The thin solid line
  represents the resolution limit of the simulation with respect to
  the number of halos. }
\end{figure}

\begin{figure}
\includegraphics[width=8cm]{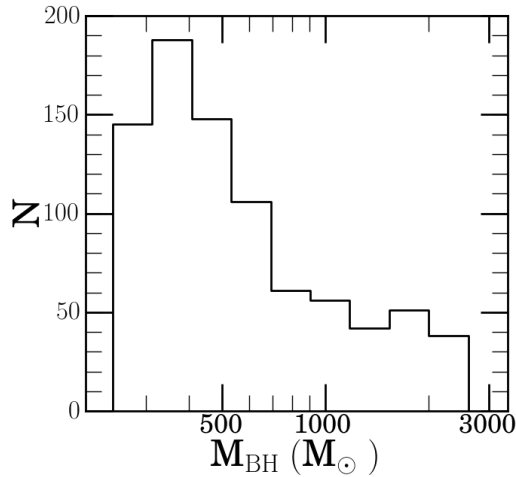}
\includegraphics[width=8cm]{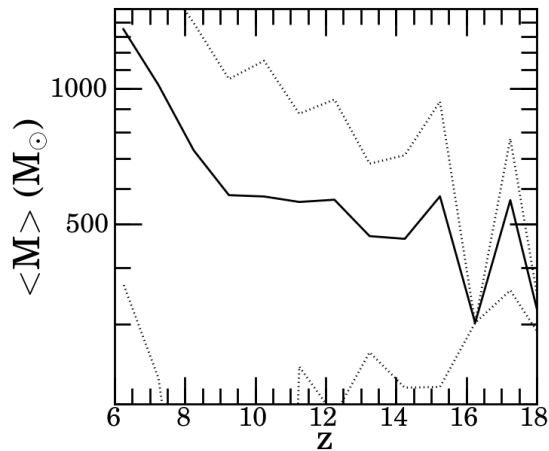}
\label{fig:mbhref}
\caption{Top panel: mass function of seed BHs formed via stellar
  dynamics in NCs, for our reference model. Bottom panel: Mean mass as
  a function of redshift $z$ for the same formation channel. Dotted
  lines denote the dispersion at $1-\sigma$ level.}
\end{figure}

\begin{figure}\label{fig:rhoref}
\includegraphics[width=8cm]{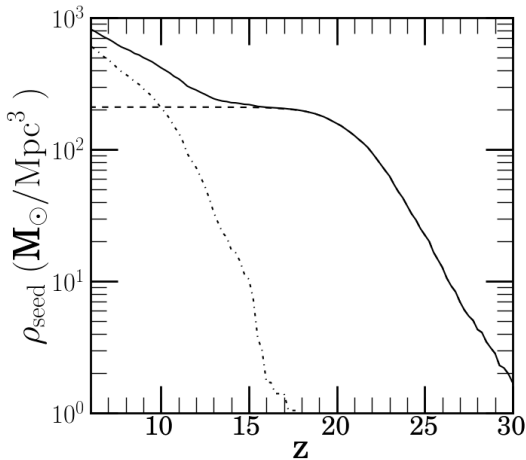}
\caption{Co-moving seed densities versus redshift $z$. Solid line:
  total co-moving BH seed density.  Dashed line: $\rho_{\rm seed}$
  versus $z$ for BH seeds formed as relics of PopIII stars more
  massive that $260\,\msun$.  Dotted line: $\rho_{\rm seed}$ as a
  function of $z$, for BHs formed via dynamical instabilities in
  NCs. }
\end{figure}

We here discuss a model in which we adopt a PopIII IMF in the range
10-300 $\msun$, an inflow efficiency $\eta=0.3$ and $Q_c=2$. The
$Z_{\rm crit}-n_{\rm crit}$ prescription adopted here is the same as
in PaperI, giving a minimum $Z_{\rm crit}\sim 10^{-5}\, Z_{\odot}$.

Figure 2 shows the fraction of virialised halos that form a seed BH,
at a given redshift, either via the dynamical (dashed line) or PopIII
(solid line) channel. The thin solid line represents our resolution
limit on the number of halo (i.e. it is one over the total number of
halos at that $z$).  As already found in previous studies BHs,
remnants of PopIII stars, start forming early on ($z \sim 30$, Yoshida
et al. 2003, Maio et al. 2010,2011, Ciardi \& Ferrara 2005). They
become increasingly rarer with cosmic time as a result of chemical and
radiative feedback\footnote{Note however that in a study by Bellovary
  et al. 2011, a second peak of BH formation is found at later
  times. They notice that this could be due to the presence of metal
  enrichment inhomogeneity within larger halos. Pockets of still metal
  free gas clouds within the larger halos, can provide suitable sites
  for PopIII formation. This effect is not visible in our simulation,
  due to the assumption of efficient mixing of metals within a single
  halo.}. In our simulations their formation stops around $z\sim
10-15$. BHs from stellar instabilities in NCs form early after the
death of the first PopIII stars. They initially are rare events:
suitable halos are only a sub-sample of those systems that previously
hosted a PopIII star. With time, metal enrichment spreads out in a
larger number of systems. The fraction of halos that form a seed BH,
at a given redshift, via the dynamical channel grows between redshift
20 and 10 from $f_{\rm BH}\sim 10^{-4}$ to $f_{\rm BH}\sim 0.002$ and
then declines. Note that the two BH formation channels considered in
this paper overlap in time only for a brief epoch.

Figure 3 (upper panel) shows the mass function of seeds formed at
$z>6$ through the dynamical channel alone. $M_{\rm BH}$ spreads in the range 300-3000 $\msun$. Most BHs
have masses around a few hundred $\msun$, with a tail that extends up
to 3000 $\msun$. Figure 3 (lower panel) shows mean BH seed masses
$\langle M_{\rm BH} \rangle$ versus redshift.  $\langle M_{\rm BH}
\rangle$ goes from 300 $\msun$ at $z\sim 15-20$ to $10^3 \msun$ at
$z=6$. Heavier BH seeds form at lower redshift when halos become
increasingly heavier, retaining more gas after PopIII explosions. In
these deeper potential wells, unstable discs 
build up more easily due to the higher gas content. Stronger inflows can
develop due to the higher gas densities, thus allowing for higher
$M_{\rm BH}$.

Figure 4 shows co-moving mass densities of seeds, $\rho_{\rm seed}$,
in our reference model. The dashed (dotted) line corresponds to the
PopIII (dynamical) channel, the solid line corresponds to the total
seed mass density. The PopIII channel dominates $\rho_{\rm seed}$ at
early time, from redshift $z\sim 30$ down to $z\sim 15$.  The
dynamical channel becomes important at later times ($z\sim 13$) due to
the increasing effects of the radiative and chemical feedbacks.  The
density, $\rho_{\rm seed}$, continues to increase up to the minimum
redshift of the simulation ($z=6$).  The lower number of PopIII stars
that can form (and consequently the lower number of relic BHs) at
$z\lsim20$ causes the flattening of $\rho_{\rm seed}$ around $z\sim
20$, which is followed by the rising contribution of the dynamical
channel.

\subsection{Where do BHs form and reside?}

In order for a BH seed to form, specific conditions need to be
fulfilled in the housing halo. These were discussed in Section 2 and
3.  Once formed, BHs are implanted and redistributed inside halos via
mergers. The population of BH-hosting halos is therefore shaped both
by the formation mechanisms and the hierarchical merging process. To
which extent each of these two phenomena is more relevant depends on
both halo properties and redshift.

In the local Universe, BHs are known to reside in the most massive
galaxies, their occupation fraction being equal to 1 for the most
massive systems. Figure 5 (panel a) shows the fraction $F_{\rm BH}$
of halos hosting a BH at $z=20,15,10,6$ (curves from left to right) as
a function of the halo mass $M_{\rm h}$. The occupation fraction
increases with increasing $M_{\rm h}$ in a self-similar way: as
redshift decreases the shape of $F_{\rm BH}(M_{\rm h})$ remains the
same, but shifts towards larger masses.  Above a characteristic mass
(that depends on $z$) $F_{\rm BH}=1$. This shift is due to two
effects: (i) as redshift decreases, smaller mass halos are
progressively less effective in forming a BH seed: smaller mass
systems are more susceptible to feedback effects that become
progressively more important as redshift decreases; (ii) halos that
contain BHs merge with others, shifting their mass towards the higher
mass range.

\begin{figure}\label{fig:disc}
\includegraphics[width=8cm]{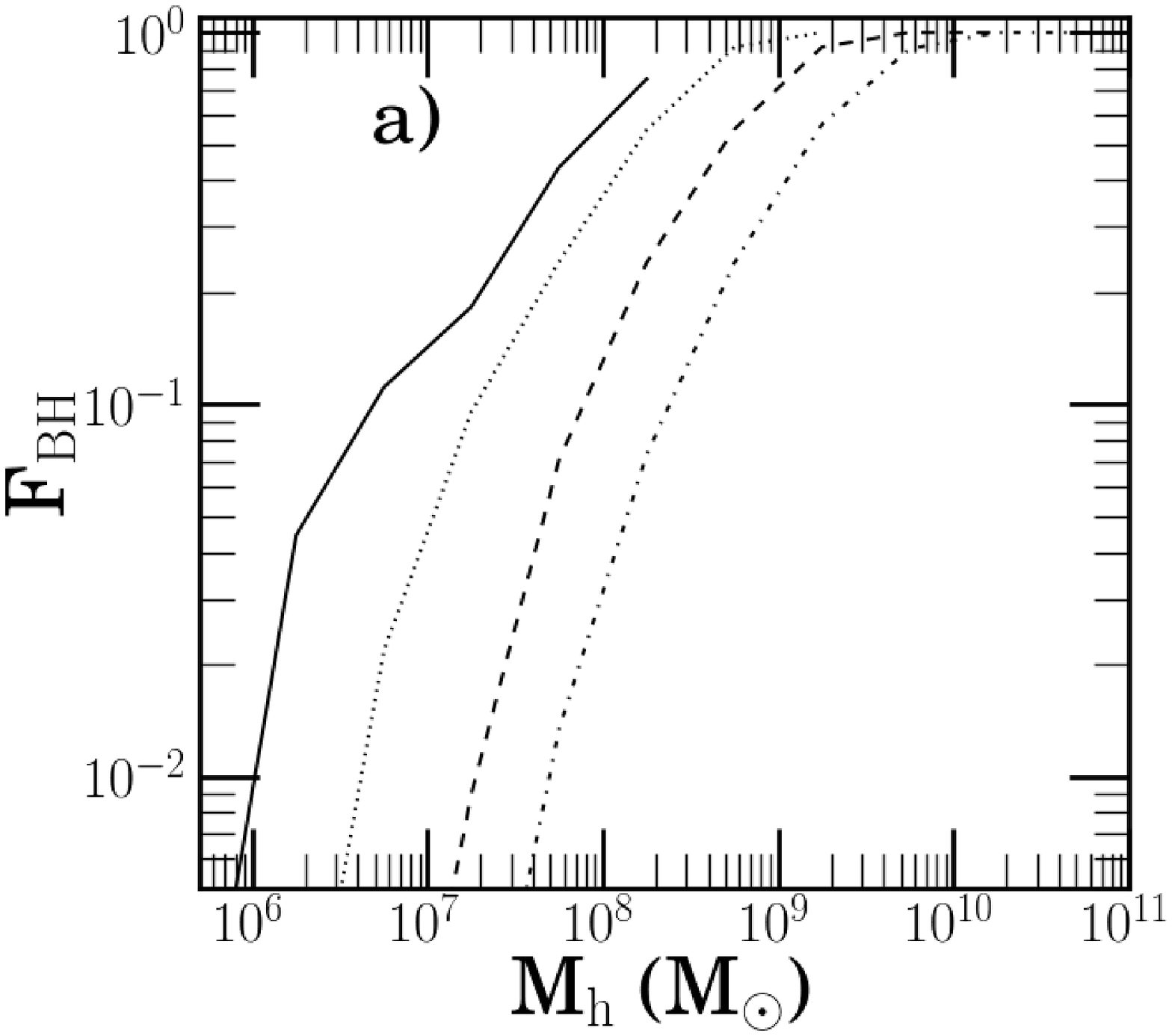}
\includegraphics[width=8cm]{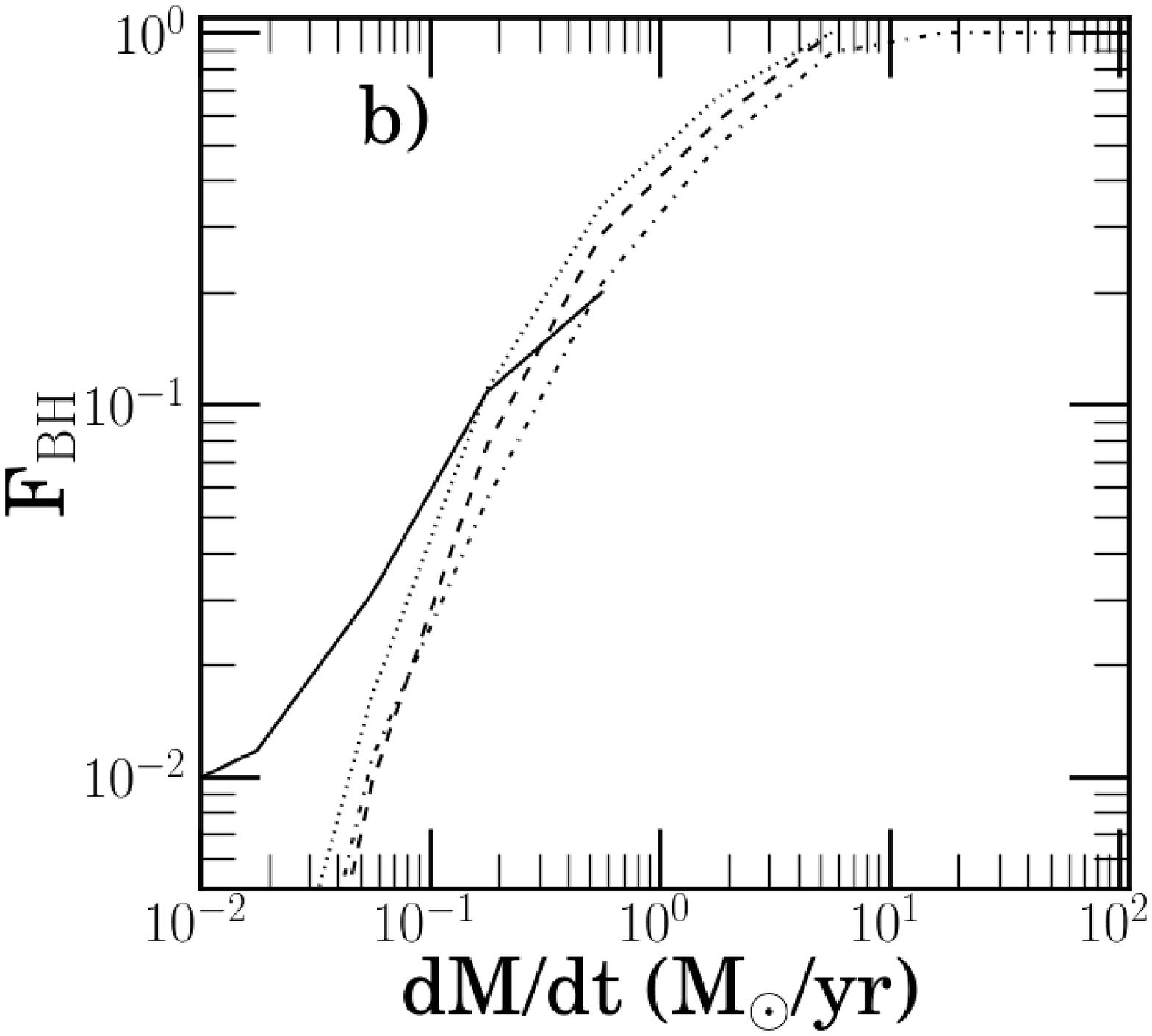}
\includegraphics[width=8cm]{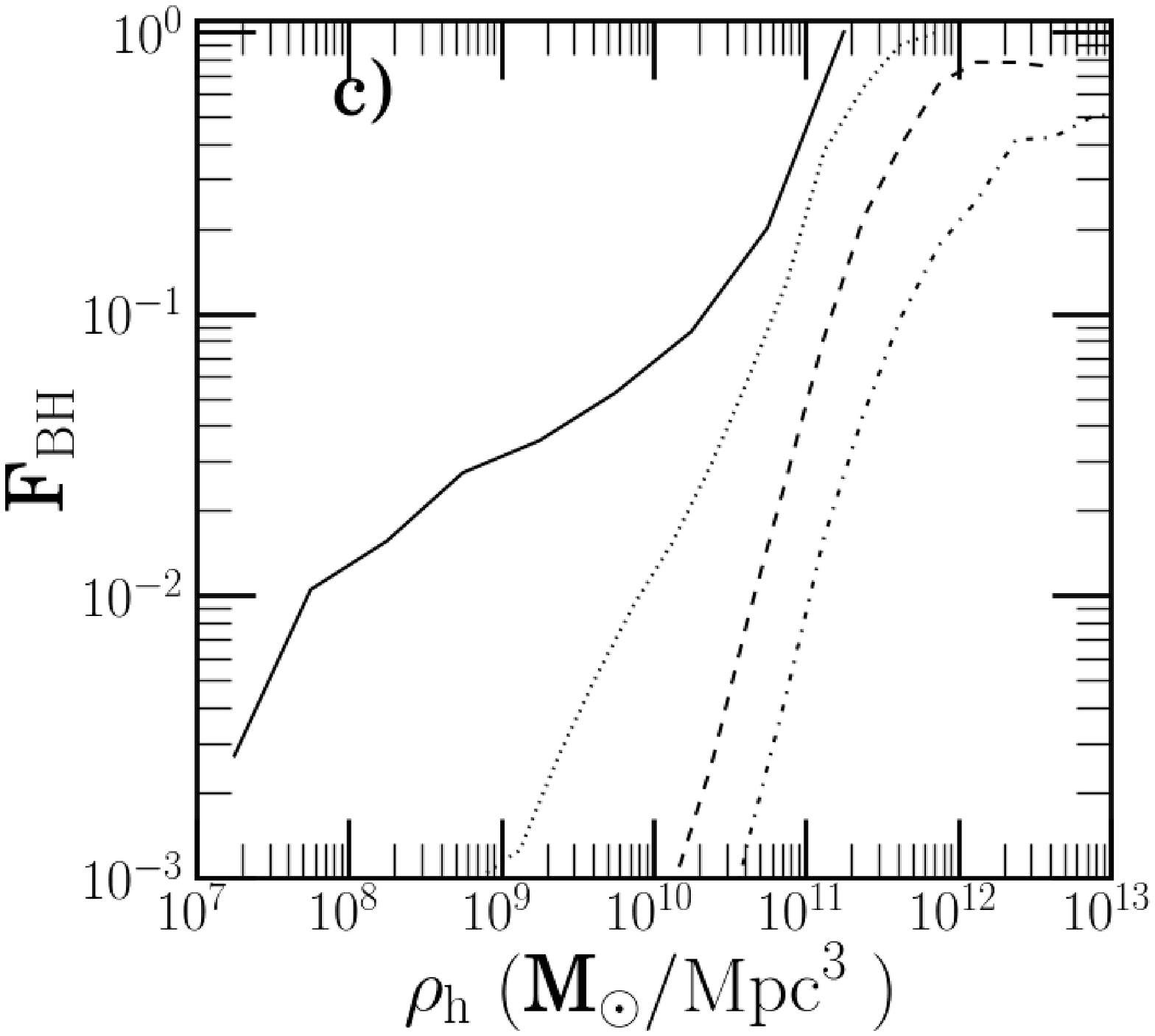}
\caption{Occupation fraction of BHs as a function of halo masses
  $M_{\rm h}$ (panel a), growth rate (panel b) and surrounding halo
  density (panel c). Solid, dotted, dashed and dot-dashed lines
  correspond to z=25, 20, 15 and 10, respectively.}
\end{figure}

Figure 5 (panel b) shows $F_{\rm BH}$ as a function of the halo mass
growth rate ${\rm d}M_{\rm h}/{\rm d}{t}$. Here ${\rm d}M_{\rm h}/{\rm
  d}{t}$ is the mean value calculated along the all halo
lifetime. Solid, dotted, dashed and dot-dashed lines correspond to
$z=20,15,10$ and $6$, respectively.  BHs reside in those faster
growing systems, that are also the more massive. This population of
hosts evolve in time, reaching higher masses and thus ${\rm d}M_{\rm
  h}/{\rm d}{t}$; $F_{\rm BH}$ shifts accordingly maintaining the same
shape.

The latter panel in Figure 5 (panel c) shows $F_{\rm BH}$ as a
function of halo density again at z=20,15,10 and 6 (solid, dotted,
dashed and dot-dashed lines, respectively).  With halo density,
$\rho_h$, we here refer to the environmental density of surrounding
the halos. We use a mass-weighted indicator, that accounts for the
number and mass of the neighbouring halos.  To calculate $\rho_h$ we
follow the following procedure: we first identify the sphere centred
on the halo, that contains its closest 32 neighbours. We then
calculate the dark matter density within that sphere and assign this
value as $\rho_h$.  In this case no self-similar trend is found in the
behaviour of $F_{\rm BH}$. The probability of hosting a BH is higher
for halos residing in denser environments. This is consistent with the
fact that the more massive halos usually reside in a crowded
environment, where they can also grow faster. With decreasing redshift
this effect becomes progressively more relevant and $F_{\rm BH}$
steepens considerably for high $\rho_h$.

\begin{figure}\label{fig:disc}
\includegraphics[width=8cm]{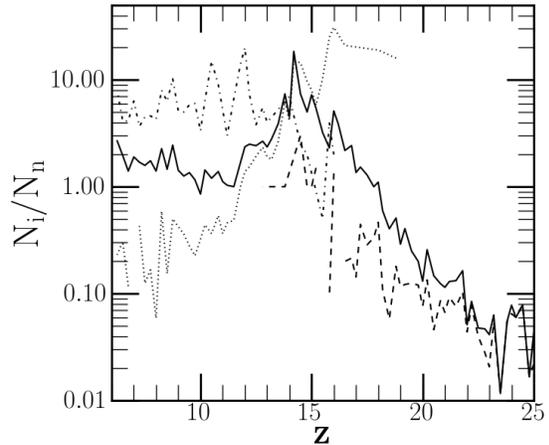}
\caption{Ratio of the number of inherited (N$_i$) and native (N$_n$)
  BHs versus redshift. Solid line corresponds to N$_i$/N$_n$ calculated
  for the all halo population. Dashed, dotted and dot-dashed lines
  correspond to N$_i$/N$_n$ calculated including only halos with
  $M_{\rm h}< 10^7 \msun$, $10^7\, \msun <M_{\rm h}< 10^8 \msun$ and
  $M_{\rm h}> 10^8 \msun$, respectively.  }
\end{figure}

A halo can host a BH either because the BH formed there as {\it
  native} BH, or because it was deposited after a merger with a
secondary system ({\it inherited} BH).  Figure 6 shows the ratio
between the number of inherited (N$_i$) and native (N$_{n}$) BHs as a
function of $z$. Solid lines correspond to N$_i$/N$_n$ computed for
all halos.  N$_i$/N$_n$ depends on $M_{\rm h}$, ${\rm d}M_{\rm h}/{\rm
  d}t$ and $z$. In order to capture this dependency, we compute this
ratio selecting halos in different mass bins. Different trends can be
inferred:
\begin{itemize}
\item The number of native BHs versus $z$ is a double peaked function,
  reflecting the different formation efficiency with $z$ of the two
  formation mechanisms considered here. The number of inherited BHs,
  increases with decreasing $z$. The ratio N$_i$/N$_n$ initially
  increases, and reaches unity around $z\sim 15-20$ (solid line). At
  higher $z$ mergers do not have time enough to distribute BHs in
  sites different from their original formation place: only halos that
  form a BH, host one. In this regime $F_{\rm BH}$ reflects the
  efficiency of the formation process. At lower $z$, conversely,
  $F_{\rm BH}$ is dominated by the ability of halos to acquire and
  implant an already existing BH. We therefore confirm the result
  originally found by Menou et al. (2001, see also Volonteri et
  al. 2003). N$_i$/N$_n$ decreases again during the second peak of BH
  formation (via the dynamical channel) but always remains higher then
  one.
\item Lower mass halos (dashed line in Figure 6) have a high
  probability of hosting a native BH. On the contrary, higher mass
  halos (dot-dashed line) more commonly acquire their BHs via
  mergers. The characteristic mass at which this transition appears
  shifts toward higher $M_{\rm h}$ with decreasing $z$. This causes
  halos in the mass range $10^7-10^8\,\msun$ (dotted line) to have
  N$_i$/N$_n > 1$ at higher $z$ ($\gsim$ 12), and N$_i$/N$_n < 1$
  afterwards.
\item At a given redshift, halos with higher ${\rm d}M_{\rm h}/{\rm
  d}t$ more often acquire their central BHs through mergers. The
  transition between the two regimes is around ${\rm d}M_{\rm h}/{\rm
    d}t\sim 1-2$ M$_{\odot}$ yr$^{-1}$, almost independent of $z$.
\end{itemize}

\begin{figure}\label{fig:rates}
\includegraphics[width=8cm]{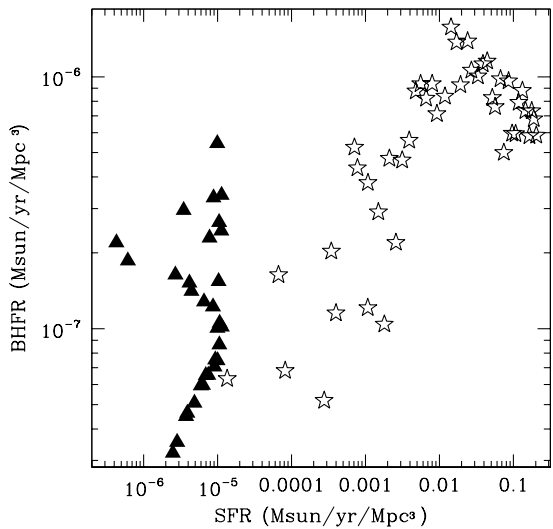}
\caption{Black hole formation rates (BHFR) versus star formation in
  our cosmological volume. Triangles refer to PopIII channel while
  stars refer to the dynamical channel. Points correspond to snapshots
  of our simulation taken at a different redshift. The $z$ interval
  considered spans between 10 and 40.}
\end{figure}

One potential caveat on the discussion above is related to the
possibility that the central BH is ejected from its host after a
merger with a secondary BH. Anisotropic gravitational wave emission
can impart to the remnant high recoil speeds (Baker et al. 2007,
Campanelli et al. 2007, Gonzales et al. 2007, Herrmann et al. 2007,
Koppitz et al. 2007, Schnittman \& Buonanno 2007). The strength of the
recoil depends on the mass ratio of the two BHs and on their spin
orientation. In our simulations we do not follow accretion onto BHs,
so we do not have information about mass ratio and spin orientation of
the merging components. To estimate, at least at first order, how
ejections can deplete the BH population, we used BH merger histories
extracted from our outputs. We impart a kick velocity at each merging
binary, at the time of merger, according to the following scheme. We
assume BHs are ejected when the kick velocity is larger than the
virial velocity of the host halo. We assume a flat distribution of
spin magnitudes and randomly oriented BH spins. We adopt this
configuration as a potential upper limit to the effect of recoils
(Volonteri et al. 2010), that could potentially lead to the largest
changes in our results. We adopt three different prescriptions to
calculate the mass ratio of the merging BHs in a binary:
\begin{itemize}
\item we keep the mass of the seed BHs without considering any
  accretion;
\item we consider $M_{\rm BH}=M_{\rm inf}$, i.e. the case in which
  all the mass channelled in the centre of the halo is accreted into
  the BH;
\item we assume a simple scaling relation to hold between $M_{\rm BH}$
  and $M_{\rm h}$ so that the BH mass ratio is equal to the mass ratio
  of their host halos.  This prescription is inspired by the $M_{\rm
    BH}-M_{\rm bulge}$ correlation (H{\"a}ring \& Rix 2004; Magorrian
  et al. 1998).
\end{itemize}

In all three cases the fraction of halos that host BHs changes only by
less then 15\% at $z=6$ (for similar studies on the effect of
ejections on BH occupation fraction see also Volonteri et al. 2010,
Schnittman 2007). Note that this is an upper limit to the actual
efficiency of ejection since in the systems considered in this paper,
we would expect the merger to happen in an environment where the two
BHs accrete gas from a gas reservoir that has a net angular
momentum. The spins of the two BHs are then expected to align with the
angular momentum of the binary (Bardeedn \& Petterson 1975, Perego et
al. 2009, Dotti et al. 2010), leading, typically, to lower recoil
velocities.

BHs arising from the dynamical channel are expected to form more
easily in gas rich structures with high inflows. In these objects,
global star formation is easily activated. As a consequence, we expect
the formation of BH seeds through the dynamical channel to be related
to the star formation rate in galaxies. Figure 7 shows the total BH
formation rate versus the star formation rate in our simulated
cosmological volume for the two channels. Different points correspond
to different simulation snapshots, taken at interval of redshift
0.1. The star formation rate increases with cosmic times. From Figure
7, it appears that our model predicts a correlation between the global
star formation rate and the formation rate of BHs only for the
dynamical channel and up to 0.1 $\msun$ yr$^{-1}$ Mpc$^{-3}$. At
higher star formation rates the correlation changes sign. This is due
to the effect of metal enrichment that increases the typical metal
content in halos. At first this increases the number of halos whose
gas has metallicity $Z_{\rm crit}<Z<10^{-3} Z_{\odot}$. As more metals
are released, the typical $Z$ of halos suitable for hosting a runaway
collision in their centre increases above our threshold for BH
formation. BH formation models that require metal free gas do not
produce this correlation (see Bellovary et al. 2011).

\subsection{Changing model parameters}

We here discuss how our results are affected by changing model
parameters, specifically how our results depend on $Q_c$, $\eta$,
$Z_{\rm crit}$ and the PopIII IMF. We change each parameter at a time,
fixing all the others. Simulations are run with $Q_c=1$, $\eta=1$,
PopIII masses between 1-300 $\msun$ and $10-600\,\msun$. We adopt the
different $Z_{\rm crit}-n_{\rm crit}$ relationship discussed in
DV09. As a reference, minimum values of $Z_{\rm crit}$ considered are
$10^{-6}\, Z_{\odot}$ and $10^{-4} \, Z_{\odot}$. We run simulations
up to $z=10$ and compare their results with our reference model at
that same $z$.

A general result of these simulations is that $\langle M_{\rm BH}
\rangle$ does not depend sensibly from the model parameters, ranging
between 600-800 $\msun$ in all our simulations.  The main effect that
changing model parameters have is on the number of BHs formed. This
consequently affects $\rho_{\rm seed}$.

\begin{itemize}
\item {\it PopIII masses:} increasing the maximum mass of PopIIIs does
  not affect our results. Allowing for a lower minimum mass increases
  considerably the number of low mass stars, given our chosen IMF.
  Metals released by these stars are usually not enough to lead to a
  strong imprint on their host halos. The number of halos where
  $Z>Z_{\rm crit}$ decreases. This decreases the number of halos
  suitable to form a BH seeds via the dynamical channel, decreasing
  $\rho_{\rm seed}$ by a factor 2-3. Note that we fix the shape of the
  IMF. A shallower slope would produce a higher fraction of high mass
  stars. These would initially produce a higher number of remnant
  BHs. At the same time the more frequent events of pair instability
  SNe and larger number of LW photons, would suppress PopIII formation
  earlier then in our models. The formation of BHs in NCs would also
  start earlier, as the number of halos with $Z>Z_{\rm crit}$ is
  increased.
\item{\it $Z_{\rm crit}$:} increasing $Z_{\rm crit}$ of 1 order of
  magnitude, decreases the metallicity range whitin which we allow BHs
  to form via runaway collisions. This reduces the resulting number of
  BH formed via this channel by a factor $\sim$ 30. At the other end,
  decreasing $Z_{\rm crit}$ of a factor 10, only increases the number
  of BH of less then a factor $\sim$ 2. A single pair instability SN
  typically increases the halo metallicity above $10^{-5} Z_\odot$, so
  that only a few halos have a metallicity between $10^{-6}-10^{-5}
  Z_\odot$.
\item{\it $\eta$:} higher $\eta$ allows for faster inflows and the mass of
  the NCs at the moment of BH formation is consequently higher. A
  population of NCs not massive enough to build up a VMS of $m_{\rm
    VMS}>260\,\msun$, exists in our reference model. For higher $\eta$
  these systems shift toward larger masses. This increases the number
  of BH formed by a factor 2. 
\item{\it $Q_c$:} for $Q_c=1$ only those halos hosting the most
  unstable discs are still prone to instability. These are the same
  systems in which larger inflows develop. The resulting NC system
  shows a depletion in its lower mass population and thus a reduction
  in the number of BHs, particularly at lower masses. This depletion
  lowers $\rho_{\rm seed}$ by a factor 2.
\end{itemize}

\section{Discussion}

We developed a model for BH seed formation either as remnants of
PopIII stars, or as a result of runaway stellar collisions in high
redshift NCs.  We devised a scheme for the evolution of the baryonic
component, illustrated in Figure 1. Halos with $Z<Z_{\rm crit}$ form
single PopIII stars, and stars more massive than 260 $\,\msun$ provide
for a remnant BH of comparable mass, after 3 Myrs from formation.
PopIII stars are the first sources of radiative, chemical and
mechanical feedbacks that affect further evolution of the gas in dark
matter halos. Once halos are polluted above $Z_{\rm crit}$ we assume
that PopII/I star formation sets in. Gas settles down in a disc
structure of given mass and angular momentum.  Inflows and star
formation compete in driving the evolution of the discs that may
become unstable.  Metal poor NCs (with $Z_{\rm crit}<Z<10^{-3}
Z_{\odot}$) formed in the central region of the disc, can provide
suitable sites for runaway collisions, leading to the formation of a
VMS and thus of a seed BH.

We explored the properties of the population of evolving halos,
combining the clustering of dark matter halos (followed with
PINOCCHIO), with our semi-analytical model. Our aim was that of
comparing the efficiency of the two mechanisms, PopIII remnants versus
dynamical collisions, in shaping the BH seed population.

PopIII stars form BH seeds already at redshift $z\sim 30$, and this is
in agreement with previous results. Their rate of formation decreases
with decreasing redshift and this channel becomes sterile after $z\sim
15-20$, due to the metal diffusion caused by SN explosions and H2
dissociation due to LW photons. BH seeds from the dynamical channel
start forming early after the rising of the first PopIII. This is due
to the rapid pollution that can develop in a halo where a PopIII
exploded.  But only at later times (i.e., around $z\lsim 15$), the
dynamical channel provides an efficient mechanism for the formation of
a larger number of seeds.  In addition, masses for seed BHs from the
dynamical channel can reach values up to $\sim 10^3\,\msun$ and are
typically higher than those left behind by PopIII stars
($200\,\msun$).  These two facts both contribute in a rapid increase
of seed mass co-moving densities. The co-moving mass density from
runaway collisions exceeds that of PopIII remnants below redshift
$z\lsim 13$.

The BH occupation fraction is sensitive to the halo mass, growth rate
and density along the redshift interval spanned in our simulation. BHs
reside in higher mass, faster growing systems, that inhabit over-dense
regions of the Universe. Halos with these characteristics are more
suitable for the formation of native BHs, and/or have higher
probabilities to inherit a BH via a merger with smaller systems.

Observations of galaxies at $z=0$ indicate that the occupation
fraction in systems more massive then $\sim 10^{11} M_{\odot}$ is
unity (Decarli et al. 2007, Gallo et al. 2008). To check that this is
the case also in our model, we analyse the merger tree histories of
subsets of our halos up to $z=0$. Halos above $10^{11}\, M_{\odot}$
all host a BH, most probably inherited already before $z=6$. Our
predicted $F_{\rm BH}$ at $z=0$ drops below 10\% at $M_{\rm h}\,\sim
10^9 \, M_{\odot}$. Below this mass BHs are very rare, and when
present they have a probability higher then 50\% of being native BHs.

\begin{figure}\label{fig:disc}
\includegraphics[width=8cm]{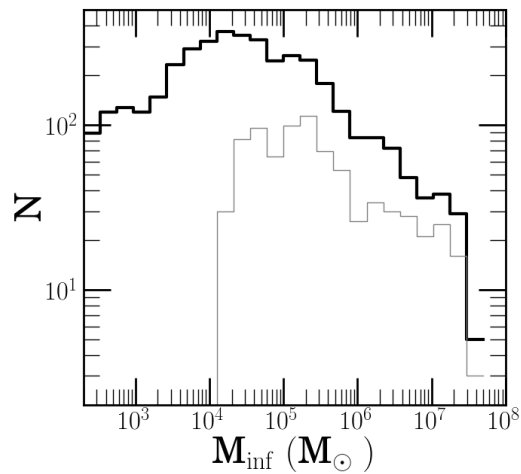}
\caption{The distribution of mass $M_{\rm inf}$ inflowing at the
  centre of halos, accumulated until redshift $z=6$ (solid
  line). The thin solid line corresponds to those systems that host a BH
  in their centre, and this represents an upper limit to the BH mass.}
\end{figure}

In our model we assumed that PopIII stars form as single objects in
any given halo. Recent studies however have shown that even metal free
gas can fragment into multiple clumps, possibly leading to the
formation of a PopIII stellar cluster (Turk et al. 2009, Stacy et
al. 2009, Prieto et al. 2011, Clark et al. 2011).  This fragmentation
process can lead to lower PopIII star masses, eventually precluding
the formation of objects with masses higher that 260 $\msun$. This
could in principle inhibit the formation of BH seeds via this
channel. We note however, that PopIII proto-clusters form as high
density systems (Clark et al. ) where collisions between stars could
still lead to the formation of a VMS (Devecchi et al. in preparation).

In this paper we ignore BH growth due to accretion of gas as we do not
take into account the competition of BHs and NCs in sharing inflowing
gas after their formation. We can infer safe upper limits for the BH
or NC masses investigating the amount of gas that flows into the
nucleus. In Figure 8 we plot (solid line) the distribution of mass
$M_{\rm inf}$, accumulated in the centre of the halo up to redshift
$z=6$.  The thin line corresponds to the same distribution inferred
for those systems that host a BH, regardless the BH is native or
inherited, and irrespective to the formation channel. Systems at $z=6$
have had time to accumulate a mass that covers a range from a few
$10^3-10^4\,\msun$, up to $10^8\,\msun$.  Note that BH seeds are
clustered in the high mass tail of the $M_{\rm inf}$ distribution.
This is suggestive that they can reach the mass range of supermassive
BHs already at redshift $z\sim 6$.  Note however, that even assuming
that all $M_{\rm inf}$ goes into growing a BH, BHs weighing billion
solar masses are not present in our simulated volume, possibly because
of to its limited size.  Very massive halos are not statistically
represented in our volumes. Given the number density of $z=6$ QSOs,
the number of BH with masses $>10^8-10^9\,\msun$ expected in our
cosmological volume is less than unity.

The novel channel of BH formation via runaway collisions in high
redshift pre-galactic discs, metal enriched by the first generation of
PopIII stars, is a promising path for the formation of seeds.  This
holds true as long as metal enrichment and diffusion is mild, i.e. as
long as the mean metallicity remains sufficiently low, such that
runaway collisions can produce a very massive star.

\section*{Acknowledgements}
This work has been supported by the Netherlands Research Council
(NWO), via grant 639.073.803 and the Netherlands Advanced School for
Astronomy (NOVA).  BD thanks Ruben Salvaterra, Francesco Haardt and
Jeroen B\'edorf for useful comments on this work. We thank the
referee, Martin Rees, for valuable suggestions.

\end{document}